\documentclass{IEEEtran4PSCC}

\ifCLASSINFOpdf
   \usepackage[pdftex]{graphicx}

\else

   \usepackage[dvips]{graphicx}

\fi

\usepackage{stfloats} 
\usepackage[cmex10]{amsmath}
\usepackage{amssymb}
\usepackage{mathtools}
\usepackage{amsthm}
\usepackage[ruled,vlined]{algorithm2e}
\usepackage{xcolor}

\usepackage{cite}
\newcommand{\citeex}[2]{[\citen{#1}, #2]}

\makeatletter
\let\old@ps@headings\ps@headings
\let\old@ps@IEEEtitlepagestyle\ps@IEEEtitlepagestyle
\def\psccfooter#1{%
    \def\ps@headings{%
        \old@ps@headings%
        \def\@oddfoot{\strut\hfill#1\hfill\strut}%
        \def\@evenfoot{\strut\hfill#1\hfill\strut}%
    }%
    \def\ps@IEEEtitlepagestyle{%
        \old@ps@IEEEtitlepagestyle%
        \def\@oddfoot{\strut\hfill#1\hfill\strut}%
        \def\@evenfoot{\strut\hfill#1\hfill\strut}%
    }%
    \ps@headings%
}
\makeatother

\psccfooter{%
        \parbox{\textwidth}{\hrulefill \\ \small{24th Power Systems Computation Conference} \hfill \begin{minipage}{0.2\textwidth}\centering \vspace*{4pt} \includegraphics[scale=0.06]{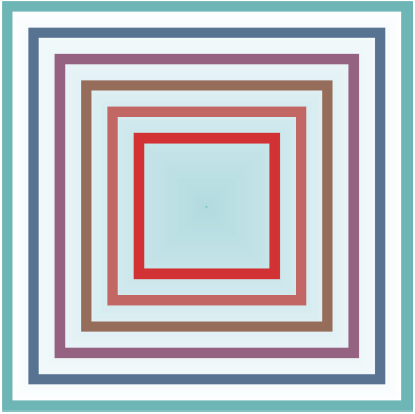}\\\small{PSCC 2026} \end{minipage} \hfill \small{Limassol, Cyprus --- June 8-12, 2026}}%
}

\begin{document}

\title{Model-Free Power System Stability Enhancement with Dissipativity-Based Neural Control}

\author{\IEEEauthorblockN{Yifei Wang\IEEEauthorrefmark{1}, Han Wang\IEEEauthorrefmark{1}, Kehao Zhuang\IEEEauthorrefmark{2}, Keith Moffat\IEEEauthorrefmark{3} and Florian D\"orfler\IEEEauthorrefmark{1}}
\IEEEauthorblockA{\IEEEauthorrefmark{1} Automatic Control Laboratory, ETH Zurich, Zurich} 
\IEEEauthorblockA{\IEEEauthorrefmark{2} College of Electrical Engineering, Zhejiang University, Hangzhou, China}
\IEEEauthorblockA{\IEEEauthorrefmark{3} Department of Electrical and Electronic Engineering, The University of Melbourne, Melbourne, Australia}
\IEEEauthorblockA{\{wangyife, hanwang1, dorfler\}@ethz.ch, keith.moffat@unimelb.edu.au, zhuangkh@zju.edu.cn}}

\maketitle

\begin{abstract}
The integration of converter-interfaced generation introduces new transient stability challenges to modern power systems. Classical Lyapunov- and scalable passivity-based approaches typically rely on restrictive assumptions, and finding storage functions for large grids is generally considered intractable. Furthermore, most methods require an accurate grid dynamics model. To address these challenges, we propose a model-free, nonlinear, and dissipativity-based controller which, when applied to grid-connected virtual synchronous generators (VSGs), enhances power system transient stability. Using input–state data, we train neural networks to learn dissipativity-characterizing matrices that yield stabilizing controllers. Furthermore, we incorporate cost function shaping to improve the performance with respect to the user-specified objectives. Numerical results on a modified, all-VSG Kundur two-area power system validate the effectiveness of the proposed approach.
\end{abstract}


\section{Introduction}
Modern power systems are undergoing significant transformations with the increasing penetration of renewable energy \cite{milano2018foundations}. Renewable resources such as wind and solar power are interfaced with the grid through power electronics converters. These converters can be controlled to behave as virtual synchronous generators (VSGs) that emulate traditional synchronous generators (SGs) in either grid-forming or grid-following mode \cite{chen2020modelling}. Conventional SG control schemes, constrained by the physical limitations of the SG, are inflexible and can be ineffective for grid stabilization challenges such as subsynchronous oscillation damping \cite{harnefors2007analysis}. This is particularly true for low-inertia grids \cite{fang2018role}. In contrast to SGs, converters are flexible and many more advanced control structures \cite{lu2019grid,wang2023power,he2024cross, leng2025deepconverter} have been developed.

Traditionally, power system transient stability analysis and subsequent control design have been based on (or have aspired to using) Lyapunov’s direct method \cite{kundur1994power, pai1989energy}. However, analytically finding Lyapunov or energy functions is often challenging, as the derivation is often highly model-specific and can suffer from the lack of a systematic methodology \cite{anghel2013algorithmic}. Moreover, the problem remains unsolved even for simple system models if certain unrealistic assumptions, such as the lossless property of the system, are relaxed \cite{cui2022equilibrium, anghel2013algorithmic}. Studies \cite{urata2018dissipativity, nahata2020passivity, cui2022equilibrium, cui2023structured} have employed passivity theory to perform stability analysis or to design the control and the Lyapunov function simultaneously. However, restrictive assumptions are still required for passivity to hold. As the complexity of power systems increases, it is advantageous to directly investigate \textit{dissipativity}, which is a generalization of passivity \cite{hill2022dissipativity}. For example, the studies \cite{martinelli2023interconnection, nakano2025dissipativity} present dissipativity-based control of DC microgrids. Nevertheless, extensions to nonlinear, AC power system applications, such as the transient angle stability of VSGs, remain largely unexplored.

Dissipativity-based control requires the design of storage and supply rate functions, which are difficult for nonlinear power system even when the dynamics are known. Current literature mainly considers the special case of Lyapunov or energy functions, but can provide heuristics for dissipativity characterization. To overcome the difficulty of finding analytical expressions discussed previously, studies have used numerical techniques such as linear matrix inequalities \cite{vu2015lyapunov} or sum of squares programming \cite{anghel2013algorithmic} to provide more general frameworks for Lyapunov function design, but these methods are still constrained by the expressivity of the chosen parametrization. Alternatively, studies have taken advantage of neural network (NN) expressivity to learn Lyapunov functions \cite{chang2019neural} and to find larger stability regions for power systems \cite{zhao2021neural, wang2022neural,nellikkath2024physics}. However, the use of NNs to characterize dissipativity has received limited attention.

In addition to the challenge of finding dissipativity-characterizing functions, power system engineers also face model-accuracy challenges. One option is to use NNs to learn the dynamics before learning the storage function (or the Lyapunov or energy function) \cite{zhou2022neural}. However, such a two-stage learning procedure may suffer from error accumulation. This motivates our data-driven approach, which \textit{learns the dissipative properties directly from the data}. 

The main contributions of this work are as follows. We develop a novel method that directly learns the dissipativity of VSG-based power systems with properly designed neural networks, which output general, symmetric, or symmetric positive definite matrices. We introduce loss functions corresponding to violations of dissipativity and stability conditions so that the matrix-generating NNs compose a stabilizing control for the system. Using the extra degrees of freedom in the dissipativity condition, we also add cost function shaping to enhance the optimality of our control with respect to user-defined performance metrics. Finally, we conduct numerical transient stability experiments on a single converter infinite bus (SCIB) system and the modified Kunder two-area system with four VSGs, demonstrating the effectiveness of our approach.

This paper is organized as follows. Section \ref{Pre} introduces the background knowledge. Section \ref{Method} presents our dissipativity-based neural control. Section \ref{sec:VSG} demonstrates how the proposed control is used in the control of VSGs. Section \ref{Results} presents the numerical experiments. Section \ref{Conclusion} concludes the paper.

\section{Preliminaries}\label{Pre}
This section introduces preliminary knowledge that supports our control design.

\subsection{Power System Transient Stability}\label{sec:TSA}
In transient analysis of converter-interfaced power systems, it is common to ignore the faster inner-loop dynamics and to use a lower-order model, such as the following second-order swing equation in a VSG-controlled system
\begin{equation}
\left\{
\begin{aligned}
    M_i\frac{\text{d}\omega_i}{\text{d}t} &= P_{\text{ref},i}-P_{\text{e},i}(\delta)-D_i(\omega_i-1) + u_i \\
    \frac{\text{d}\delta_i}{\text{d}t}&=\omega_\text{B}(\omega_i-1)
\end{aligned}
\right.\label{Swing}
\end{equation}
for VSG at bus $i\in\{1,2,\cdots,N\}$, where $M_i$, $P_{\text{ref},i}$, $D_i$, $\omega_i$, and $\delta_i$ are the moment of inertia (in s), the power reference (p.u.), the damping coefficient (p.u.), the angular frequency (p.u.), and the relative voltage angle (rad) with respect to a synchronous reference frame, and $\omega_\text{B}$ is the base (nominal) frequency in rad/s. A control input $u_i$ can be additionally added to the frequency dynamics of the $i$-th VSG. The electrical power output of converter $i$ is
\begin{equation}
P_{\text{e},i}
= E_i^2 G_{ii}
+ \sum_{\substack{j=1,\  j\neq i}}^{N} E_i E_j\!\left[\, G_{ij}\cos(\delta_{ij})
+ B_{ij}\sin(\delta_{ij}) \,\right]\notag,
\end{equation}
where $E_i$ is the virtual electromotive force of the $i$-th VSG determined by state variables, $\delta_{ij}=\delta_i-\delta_j$ represents voltage angle difference, and $G_{ij}$ and $B_{ij}$ are the conductance and susceptance elements of the reduced network admittance matrix between VSG $i$ and $j$ after Kron reduction.

Compactly, the dynamics of the power system are described by a set of continuous-time differential equations. However, we consider discrete-time dynamics to accommodate discrete-time computational control. Assuming a constant sampling and control interval $\Delta t$ and corresponding zero-order hold inputs, the continuous dynamics can be described at sampling instants by the following difference equation:
\begin{equation}
 x_{k+1}=f(x_k,u_k)\label{DynamicalSystem},
\end{equation}
where, more generally, $x_k\in\mathcal X\subset\mathbb R^n$ and $u_k\in\mathcal{U}\subset\mathbb R^m$ are the states and inputs defined over a compact set $\mathcal{X}\times\mathcal U$; the nonnegative integer $k$ is the discrete time step; $f:\mathcal X\times\mathcal{U}\to \mathbb R^n$ is locally Lipschitz continuous on $\mathcal X\times\mathcal{U}$. When $u$ becomes a function of $x$ through a locally Lipschitz continuous feedback control law $u_k=\pi(x_k),\pi:\mathcal{X}\to\mathcal{U}_{\pi}\subseteq\mathcal U$, the closed-loop system becomes autonomous:
\begin{equation}
    x_{k+1}=f(x_k,\pi(x_k))=f_\pi(x_k) \label{Closed}.
\end{equation}

The local asymptotic stability of \eqref{Closed} is defined as follows.

\textbf{Definition 1}: Suppose $x^*$ is an equilibrium point of \eqref{Closed}, i.e. $f_\pi(x^*)=0$, then the system is \textit{stable} with respect to $x^*$ if $\forall\varepsilon>0, \exists$ $\eta$ $>0,$ such that $\Vert x_0-x^*\Vert<$  $\eta$$\Rightarrow\forall k\geq0,\Vert x_k-x^*\Vert<\varepsilon$. Further, it is \textit{(locally) asymptotically stable} if $\exists$ $\eta'$ $>0$, such that $\Vert x_0-x^*\Vert<$ $\eta'$ $\Rightarrow\lim_{k\to \infty}\Vert x_k-x^*\Vert=0$. The set that contains all $x'$ such that $x_0=x'\Rightarrow\lim_{k\to\infty}x_k=x^*$ is termed as the \textit{region of attraction} (RoA) of $x^*$.

Classical transient stability analysis in power systems seeks to determine whether the system states, after a large disturbance such as a fault, are in the RoA of the (locally) stable post-fault equilibrium. With appropriate control, the RoA of the post-fault equilibrium can be enlarged and the transient stability can be improved.

\subsection{Lyapunov Function and Dissipativity}\label{sec:Disp}
Without loss of generality, we assume $f(0,0)=0$ in \eqref{DynamicalSystem} and $\pi(0)=0$, which leads to $f_\pi(0)=0$ in \eqref{Closed}. The following Theorem establishes the asymptotic stability of system \eqref{Closed} with respect to $x^*=0$.

\textbf{Theorem 1} (\citeex{khalil2002nonlinear}{Exercise 4.63}): The equilibrium $x^*=0$ of \eqref{Closed} is asymptotically stable if there exists a compact and non-empty set $D\subseteq\mathcal X$ that includes the origin, and there is a function $V:D\to \mathbb R$, termed \textit{Lyapunov function}, that satisfies
\begin{equation}
\begin{aligned}
&V(x_k)>0,\quad x_k\in D\ \backslash\ \{0\},\quad V(0)=0,\\
&V(x_{k+1})-V(x_k)<0,\quad \forall x_k\in D\ \backslash\ \{0\}.
\end{aligned}\label{Lyapunov}
\end{equation}

Theorem 1 is used to analyze the stability of autonomous systems, while in the analysis of open-loop systems with external inputs, dissipativity has been proposed as a generalization of Lyapunov stability.

\textbf{Definition 2} (Dissipativity): System \eqref{DynamicalSystem} is \textit{locally dissipative with respect to a supply rate function} $s(\cdot,\cdot):\mathcal X\times\mathcal{U}\to\mathbb R$ if there exist a \textit{storage function} $V:\mathcal X\to\mathbb R$ and compact and
non-empty sets $D_x\subseteq\mathcal X,D_u\subseteq\mathcal U$ containing the origins such that along system trajectories
\begin{subequations} \label{Dissipativity}
\begin{align}
&V(x_k)\geq 0,\quad x_k \in D_x \setminus \{0\},\quad V(0) = 0, \label{diss_psd} \\
&V(x_{k+1})-V(x_k) \leq s(x_k, u_k),\quad \forall x_k \in D_x,\ u_k \in D_u. \label{diss_comp}
\end{align}
\end{subequations}

The left-hand side of \eqref{diss_comp} can be interpreted as the rate of change of the stored energy of the system, which should be smaller than the energy supply rate on the right-hand side if the system is dissipative with respect to the supply rate.

In addition to designing arbitrary controllers that maintain the stability of the operating equilibrium and further enhance the transient stability of the power system, further performance objectives, such as minimizing frequency deviations during transients, are often considered. Therefore, in this paper, we investigate the \textit{optimal stabilizing control problem}, which encodes asymptotic stability as a constraint and is described as follows: 
\begin{equation}
\begin{aligned}
      &{\min}_{\{u_k\}_{k=0}^\infty} &&\sum_{k=0}^{\infty}l(x_k,u_k)\\
    & \text{subject to } && x_{k+1}=f(x_k,u_k), \ x_0 \text{ given}\\
    & &&\lim_{k\to\infty}x_k=0.
\end{aligned}\label{eq:optimal}
\end{equation}
where $l(\cdot,\cdot): \mathcal X\times \mathcal U\to\mathbb R_{+}$ is a nonnegative real cost function. A solution to \eqref{eq:optimal} is referred to as an \textit{optimal stabilizing control}.

\section{Methodology}\label{Method}

In this section, we present our data-driven method for designing neural controllers for power systems with unknown dynamics. We first propose a feedback control with state-dependent gain for a category of dissipative systems and establish a sufficient condition for the control to be an optimal stabilizing one with respect to a cost function $l_\text{d}$ determined by the storage and supply rate functions. We then demonstrate how to use NNs to find storage and supply rate functions that characterize the dissipativity of unknown power systems, so that we can synthesize a stabilizing controller. Additionally, by finding storage and supply rate functions that make $l_\text{d}$ close to a user-defined cost function, the synthesized controller is close to optimal with respect to the user-defined cost function.

\subsection{Dissipativity-Based Stabilizing Control}
Studies have proposed stabilizing feedback controllers for dissipative systems. Inspired by existing static feedback controls \cite{madeira2021necessary, lima2024qsr} and continuous-time dynamic feedback control \cite{wang2025learning}, we propose a dynamic feedback control for discrete-time settings as required by the data-driven context. 

Without loss of generality, we assume $f(0,0)=0$ in the following analysis. By comparing the dissipativity conditions \eqref{diss_psd}-\eqref{diss_comp} with those stated in \eqref{Lyapunov}, it follows that if 
\begin{equation}
\pi(0)=0,\ s(x,\pi(x))<0,\ \forall x\in D/\{0\}\label {feedback_disp}
\end{equation}
and the system is dissipative with respect to $s$ with a positive definite storage function $V$, $\pi(x)$ will stabilize the system toward the origin. To this end, we have the following theorem.

\textbf{Theorem 2}: Suppose that system \eqref{DynamicalSystem} is dissipative on $D_x=\mathcal X$, $D_u=\mathcal U$ with respect to generalized $(Q,S,R)$ supply rate
\begin{equation}
s(x,u)=x^TQ(x)x+2x^TS(x)u+u^TR(x)u\label{QSR}
\end{equation}
with $Q^T(\cdot)=Q(\cdot):\mathcal{X}\to\mathbb R^{n\times n}$, $S(\cdot):\mathcal{X}\to\mathbb R^{n\times m}$, $0\prec R(\cdot):\mathcal{X}\to\mathbb R^{m\times m}$, and a positive definite storage function $V(x)$. If
\begin{equation}
    \Phi(x) \coloneqq S(x)R^{-1}(x)S^T(x)-Q(x)\succ0,\forall x\in\mathcal{X}\label{delta}
\end{equation}
holds, and the range of the following feedback control with state-dependent gain
\begin{equation}
    \pi(x)=-R^{-1}(x)S^T(x)x\label{control}
\end{equation}
is a subset of $\mathcal U$, then \eqref{feedback_disp} always holds with control \eqref{control} and supply rate \eqref{QSR}. Moreover, \eqref{control} is a stabilizing control.

\textbf{Proof:} The proof is given in Appendix~\ref{Apdx:T2}.

Theorem 2 provides us with insights on how to design stabilizing controls based on dissipativity. A natural interpretation is that the control should supply as little energy as possible to the system. To see this, consider constant $Q$, $S$, and $R$, then the supply rate becomes $u^TRu+2x^TSu+x^TQx$, a quadratic function of $u$. With $R$ positive definite, the function is minimized when $u=-R^{-1}S^Tx$ and the minimum value is exactly $-x^T\Phi(x) x$. Therefore, we obtain a low upper bound $-x^T\Phi(x) x$ for the increase in the storage function $V$ in one time step. If additionally such an upper bound is negative definite, which can be governed by the stronger condition $\Phi(x)\succ0$,  $V$ can only be strictly decreasing unless it reaches $V=0$, bringing the system to equilibrium.

The minimization of the supply rate function implies the infinite-horizon optimality of the proposed dissipativity-based control, as shown in the following theorem.

\textbf{Theorem 3}: Suppose that system \eqref{DynamicalSystem} satisfies all the conditions of Theorem 2, then the control \eqref{control} is the solution to the following optimal control problem:
\begin{equation}
\begin{aligned}
    &{\min}_{\{u_k\}_{k=0}^\infty} &&\sum_{k=0}^{\infty}\left[\tilde l(x_k,u_k)+x_k^T\Phi(x_k)x_k\right]\\
    &\text{subject to} &&x_{k+1}=f(x_k,u_k), \ x_0 \text{ given}\\
    &&&\lim_{k\to\infty}x_k=0,
\end{aligned}\label{eq:optimal_disp}
\end{equation}
where $\tilde l(x_k,u_k)=-[V(x_{k+1})-V(x_{k})]+ s(x_k,u_k)$ is a nonnegative cost function, $V$ is the storage function, and $x_k^T\Phi(x_k)x_k$ is a positive definite function of $x_k$.

\textbf{Proof:} The proof is given in Appendix~\ref{Apdx:T3}.

Although \eqref{eq:optimal_disp} involves the system dynamics $x_{k+1}=f(x_k,u_k)$, this merely reflects that the costs are evaluated along system trajectories. Theorem 3 emphasizes the inherent optimality of \eqref{control} consistent with the dissipative property characterized by any pair of valid storage $V$ and generalized $(Q,S,R)$ supply rate functions, without relying on explicit knowledge of the dynamic model.

\subsection{The Matrix Neural Networks}
In order to generate a stabilizing control according to Theorem 2, the functions $V(x)$, $Q(x)$, $S(x)$, and $R(x)$ satisfying the conditions of the Theorem are required. Even for a power system with known dynamics, finding such functions would be challenging. Furthermore, an accurate system model is not always available. Therefore, we leverage data and the function approximation ability of NNs to directly learn functions that satisfy conditions in Theorem 2.  

Certain conditions can be satisfied by appropriate NN architectures, while others are results of NN training driven by corresponding loss functions.
One such NN architecture is a ``matrix NN''---an NN that outputs a matrix, and we define the matrix NN's ``dimension'' as the dimension of its output matrix.

For simplicity, we use relative states with respect to the equilibrium of the specific post-fault system in the concerned transient stability case as the input to the NNs, and the origin in the new coordinates therefore becomes an equilibrium point. Thus, we assume that there is sufficient information for steady-state analysis of the system to determine equilibrium points, while not assuming that an accurate dynamic model is available.

\begin{figure*}[!ht]
 \vspace{-2mm}
\centerline{\includegraphics[width=0.9\linewidth]{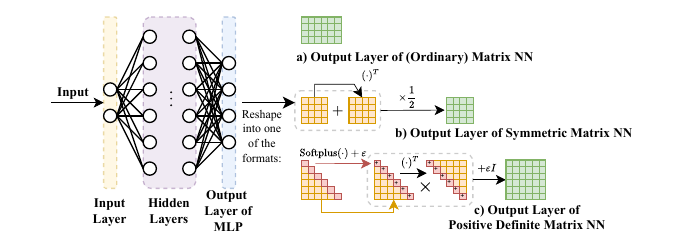}}
	\vspace{-5mm}
        \caption{The architectures of three types of matrix NNs. For simplicity, we only draw one MLP schematic. Different matrix NNs do not share the same MLP.}
        \label{MatNN}
	\vspace{-5mm}
\end{figure*}

To begin with, we have the following parameterizations:
\begin{itemize}
    \item A positive definite storage function $V(x;\theta_W)=x^TW(x;\theta_W)x$ with an $n\times n$ positive definite matrix NN $W(x;\theta_W)$;
    \item an $n\times n$ symmetric matrix NN $Q(x;\theta_Q)$;
    \item an $n\times m$ matrix NN $S(x;\theta_S)$;
    \item an $m\times m$ positive definite matrix NN $R^{-1}(x;\theta_{R^{-1}})$ whose inverse matrix NN is  $R(x;\theta_{R^{-1}})\coloneqq[R^{-1}(x;\theta_{R^{-1}})]^{-1}$; we directly parameterize $R^{-1}$ to avoid matrix inversion in \eqref{control}, thus improving efficiency in real-time application; the inverse formulation does not influence the learning stage, as both $R(x)$ and its inversion are always needed, as can be seen in the upcoming Section \ref{sec:lossfn}.
\end{itemize}
where $\theta$ represents the NN parameters. We use three types of matrix NNs that are built as follows:

\begin{itemize}
    \item An (ordinary) matrix NN of dimension $p\times q$ is a multilayer perceptron (MLP) with $n$ input units and $pq$ output units, whose output is further reshaped into a $p\times q$ matrix;
    \item A symmetric matrix NN of dimension $p\times p$ is obtained by averaging the output of an ordinary $p\times p$ matrix NN and its transpose. The transposition can be performed with standard functions such as \textit{transpose()} in PyTorch;

    \item We construct a $p\times p$ positive definite matrix NN from an MLP with $n$ input units and $p(p+1)/2$ output units. We reshape the MLP output as a lower triangular matrix $L$, pass the diagonal elements to the softplus function Softplus$(x)=\ln(1+\exp (x))>0$, add a small $\varepsilon>0$ to them, and keep the non-diagonal entries to obtain $\tilde L$, compute $\tilde L\tilde L^T$, and finally add another small $\varepsilon>0$ to the diagonal elements. The small $\varepsilon>0$ slightly restricts the approximation ability of the NN, but is beneficial in terms of numerical stability. Additionally, we allow this type of NN to output its inverse, as seen in the case of $R^{-1}(x;\theta_{R^{-1}})$.
\end{itemize}

Figure \ref{MatNN} summarizes the architectures of the matrix NNs we use.

\subsection{Loss Function Design}\label{sec:lossfn}
After building the matrix NNs, we define loss functions that penalize the violations of conditions of Theorem 2. 

According to the dissipativity condition \eqref{diss_comp} with supply rate \eqref{QSR}, we define the following loss function that penalizes the violation of the dissipativity conditions: 

\begin{equation}
\begin{aligned}
    &\mathcal L_{\text{d}}(\theta, \mathcal B)\\
    \coloneqq& \max_{d_k\in\mathcal B}\text{Softplus}\left(\varepsilon_{\text{d}}+x_{k+1}^TW(x_{k+1};\theta_W)x_{k+1}\right.\\
    &\qquad-x_k^TW(x_k;\theta_W)x_k-x_{k}^TQ(x_k;\theta_Q)x_k\\
    &\left.\qquad-2x_k^TS(x_k;\theta_S)u_k-u_kR(x_k;\theta_{R^{-1}})u_k\right)\\
    \coloneqq& \max_{d_k\in\mathcal B}\text{Softplus}(\varepsilon_{\text{d}}+(*)),
\end{aligned}
\label{loss_d}
\end{equation}
where $\theta$ refers to all the NN parameters, $(*)$ is the violation of the dissipativity condition, $\varepsilon_{\text{d}} > 0$ is a small margin that encourages the violation to be smaller and provides more safety, and $\mathcal{B}$ is a batch of data tuples $d_k=(x_k,x_{k+1},u_k)$. We use the softplus function because it is smooth and its gradient does not disappear when the input is non-positive. We use the maximum loss within a batch instead of the usual average loss because we want a conservative result where the largest violation should be non-positive.

\begin{figure*}[!ht]
 \vspace{-2mm}
\centerline{\includegraphics[width=0.9\linewidth]{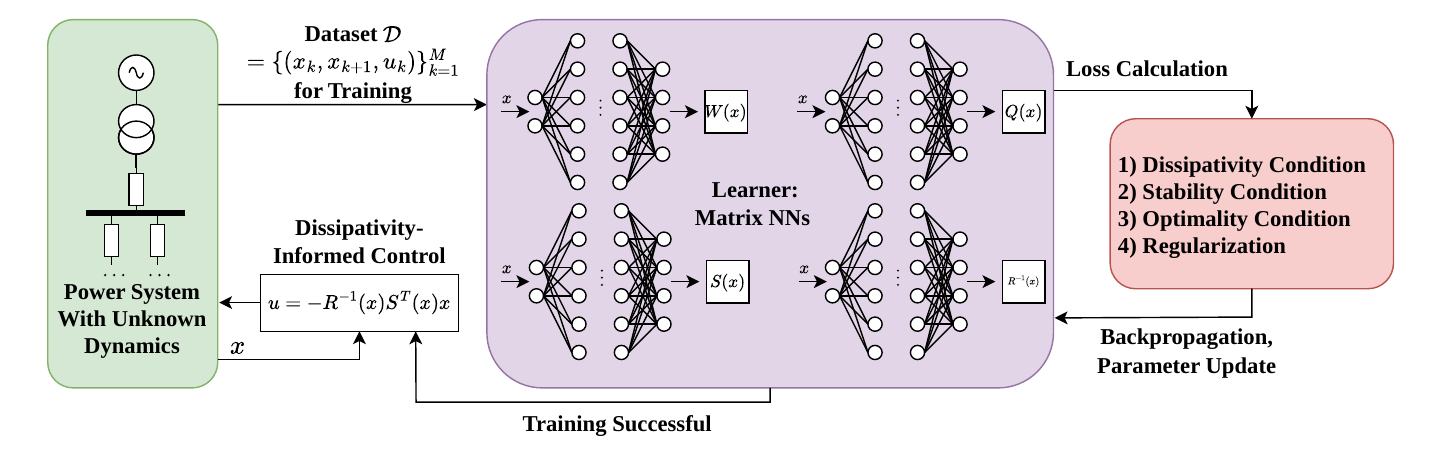}}
	\vspace{-2mm}
        \caption{An overview of our algorithm, which generates a stabilizing control after training NNs that can characterize dissipativity.}
        \label{full}
	\vspace{-4mm}
\end{figure*}

Next, the loss corresponding to stability condition \eqref{delta} is:
\begin{equation}
    \mathcal L_\Phi(\theta;\mathcal{B})\coloneqq\max_{d_k\in\mathcal B}\text{Softplus}(\varepsilon_{\Phi}-\min\text{eig}(\Phi(x_k;\theta))), \label{loss_delta}
\end{equation}
where $\varepsilon_\Phi>0$ is again a small margin that encourages larger minimum eigenvalues, and $\min \text{eig}(\cdot)$ computes the smallest eigenvalue of the input. Therefore, \eqref{loss_delta} punishes the smallest negative eigenvalue of $\Phi(x_k)$, driving it positive definite.

To utilize Theorem 3 and optimize long-term performance on top of stability, we compare \eqref{eq:optimal} and \eqref{eq:optimal_disp} and observe that if $\tilde l(x,u)+x^T\Phi(x)x$ is equal to a user-defined cost $l(x,u)$, control \eqref{control} will be optimal with respect to $l(x, u)$. Therefore, we define the following cost shaping loss function to punish the difference between the user-defined cost and the cost in Theorem 3:
\begin{equation}
    \mathcal L_{\text{sp}}(\theta;\mathcal{B})=\frac{1}{\vert\mathcal B\vert}\left[\frac{l(x_k,u_k)-\left(x_k^T\Phi(x_k;\theta)x_k-(*)\right)}{\vert l(x_k,u_k)\vert+\varepsilon_{\text{sp}}}\right]^2,\label{loss_shaping}
\end{equation}
where $(*)$ is defined as in $\mathcal L_{\text{d}}$, and $l(x,u)$ is positive definite. We use the relative error and a small margin $\varepsilon_{\text{sp}}>0$ to enhance the numerical stability.

In addition, we add a regularization term $\mathcal L_r(\theta)$ that penalizes the $\ell^1$ norm of the NN parameters. This term helps prevent parameters from becoming too large, promotes sparsity, and enhances numerical stability. The total batch loss is then
\begin{equation}
\mathcal L(\theta,\mathcal B)=w_1\mathcal L_{\text{d}}(\theta,\mathcal B)+w_2\mathcal L_\Phi(\theta,\mathcal B)+w_3\mathcal L_{\text{sp}}(\theta,\mathcal B)+w_4\mathcal L_{\text{r}}(\theta), \label{loss_total}
\end{equation}
where $w_1>0,\  w_2>0,\ w_3\geq0,\  w_4\geq0$ are tunable weights.

\subsection{Learning Dissipativity with Neural Networks}
With the defined matrix NNs and the loss functions, we use the usual neural network training workflow to train our matrix NNs. To begin with, we sample states and inputs from the trajectories of the specific unknown (post-fault) system and rearrange them into a data set $\mathcal D$ in the form of $\{(x_k,x_{k+1},u_{k})\}_{k=1}^M$ for convenience, where $M$ is the total number of tuples. We set the number of epochs and divide $\mathcal D$ into batches in each epoch. For each batch, we calculate the loss \eqref{loss_total} and update the NN parameters through backpropagation. When training is complete, the NNs generate a control $\pi(x)$ according to \eqref{control}.

The effectiveness of the algorithm is dependent on the richness of the training data. Therefore, we have to implement the returned control law $\pi(x)$ to see if it indeed stabilizes the closed-loop system. Tuning of the hyperparameters is often necessary.

Figure \ref{full} summarizes our algorithm, including matrix NN training and the design of stabilizing feedback control. Algorithm \ref{algo} provides the pseudocode of the learning process.

\begin{algorithm}
\caption{Learning Dissipativity of Power System With Stability Conditions}
\label{algo}
\KwIn{data set $\mathcal D=\{(x_k,x_{k+1},u_{k})\}_{k=1}^M$, hyperparameters (number of epochs $N_\text{e}$, batch size $\vert\mathcal B\vert$, size of MLP, learning rate, etc.)}
\KwOut{Storage function $V(x)$, supply rate function $s(x,u)$, and stabilizing control $\pi(x)$}
\textbf{Initialize:}{ Parameters $\theta$ of $W(x)$, $Q(x)$, $S(x)$, $R(x)$}\\
\For{epoch $= 0, 1, \dots, N_e$}{
    Randomly shuffle the data set $\mathcal D$\\
    Divide $\mathcal D$ into batches of size $\vert\mathcal B\vert$\\ 
    \For{batch $\mathcal B$ in batches}{
    Calculate loss $\mathcal L(\theta,\mathcal{B})$ and $\nabla_\theta\mathcal L(\theta,\mathcal{B})$\\
    Update $\theta$ with Adam \cite{kingma2014adam} optimizer 
    }}
\Return{Control $\pi(x)=-R^{-1}(x)S^T(x)x$}
\end{algorithm}

\section{Dissipativity-Based Neural Control of Virtual Synchronous Generators}\label{sec:VSG}

In this section, we describe how we apply dissipativity-based neural control to VSG inverter control. We choose VSG control as it is one of the most common grid-forming architectures. It is possible that dissipativity-based neural control could be applied to other inverter control methods and synchronous machine control.

The VSG control input $u$ is an adjustment in the rate of change of the angular frequency, as in \eqref{Swing}. 
The controller is trained with discrete-time samples and thus produces discrete-time commands $u_k$.
Thus, commands $u_k$ are applied using a zero-order hold for the control/sampling interval $\Delta t$.
The state measurement $x$ consists of $\Delta\omega$ and $\Delta\delta$, the frequency and (relative) voltage angle deviations with respect to the post-fault equilibrium.
We do not consider measurement noise in this implementation.
In addition, dissipativity-based neural control requires a cost-shaping goal. For the tests in Section \ref{Results}, we used the quadratic cost 
\begin{align}
l(x,u)=1000\Vert\Delta\omega\Vert^2+\Vert\Delta\delta\Vert^2+10\Vert u\Vert^2, \label{eqn:VSGcontrolCost}
\end{align}
where $\Vert\cdot\Vert$ is the 2-norm. 

We consider here a centralized neural controller that generates the control input for all VSGs in the system. Thus, the centralized neural controller requires measurements of the angular frequencies $\omega$ and voltage angles $\delta$ of all VSGs. 
Decentralized or distributed training and controller synthesis, noisy measurements, and using measurements that do not require knowledge of the post-fault equilibrium are directions for future work.

\section{Simulation Validation}\label{Results}
In this section, we validate our approach using numerical experiments of an SCIB system and a modified Kundur two-area system. In both cases, the nominal frequency is $60$ Hz. Both post-fault systems had a stable equilibrium, but the fault-clearing states lay outside the corresponding RoAs. For the NNs, we used GELU (Gaussian-error linear unit) as the activation function and $128$ as the output dimension of all hidden layers in every MLP. We set the cost shaping goal as in \eqref{eqn:VSGcontrolCost}. Although in the simulation we used higher-order dynamics of VSGs including low pass filter dynamics in current and voltage measurement and inner-loop PI controller dynamics, we used only $\Delta\omega$ and $\Delta\delta$ as inputs to the NNs and to the generated controller in both training and implementation stages.

\subsection{Single Converter Infinite Bus System}
\begin{figure}[!ht]
 \vspace{-2mm}
\centerline{\includegraphics[width=0.9\linewidth]{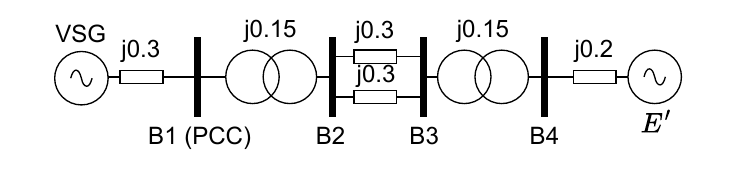}}
	\vspace{0mm}
        \caption{Schematic of the SCIB system.}
        \label{smib}
	\vspace{-3mm}
\end{figure}
As seen in Figure \ref{smib}, we considered a second-order VSG connected to an SG. The inertia of the SG was set to infinite, so the constant internal electromotive force $E'$ acted as an infinite bus. The virtual inertia of the VSG was $4$s, the damping coefficient was $5$p.u., and the virtual impedance between the virtual electromotive force and the point of common coupling (PCC) was j$0.3$p.u. The system was initially set with nominal voltage at B1 and B4, and the voltage of B1 lagged behind B4 $0.3$rad; $E'$ and the power set point of VSG were determined accordingly. 

\begin{figure}[!ht]
 \vspace{-3mm}
\centerline{\includegraphics[width=0.8\linewidth]{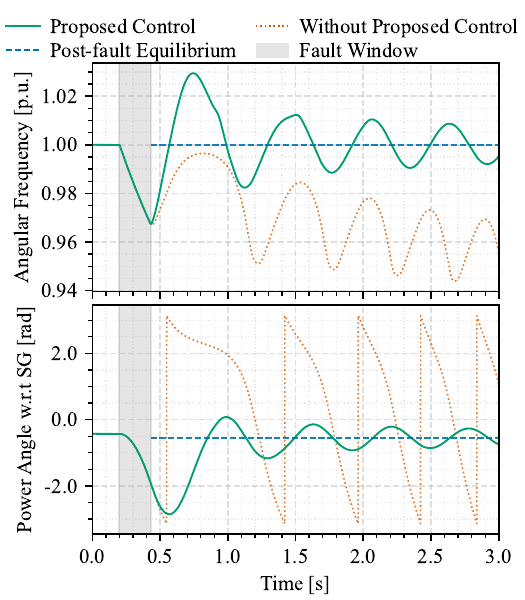}}
	\vspace{0mm}
        \caption{The effect of proposed control in the SCIB system.}
        \label{smib_ctrl}
	\vspace{-3mm}
\end{figure}

To train the matrix NNs, we first generated a data set containing $(x_k,x_{k+1},u_k)$ tuples with fixed sampling and control interval $\Delta t=0.5$ ms. As the convergence of post-fault states is decided by the post-fault dynamics, the training data were gathered in the post-fault system, which matched Figure \ref{smib} but with one line between B2 and B3 open, as described below.

To ensure sufficient excitation, we randomly sampled $u_k$ and initial conditions from uniform distributions. We generated $2000$ short trajectories of length $1$s. After simulation, we filtered the sampled data to obtain a training data set that consisted only of data in a region of interest $\tilde{\mathcal{X}}$ where $|\Delta\omega|\leq0.1$ and $|\Delta\delta|\leq\pi$. The filtering process ensures that the training focuses on a typical operating region of the system and helps prevent unrealistic far-from-equilibrium data from distorting the gradients. We used a learning rate of $5\times10^{-4}$ and weight decay $10^{-4}$ in Adam, $(w_1,w_2,w_3,w_4)=(10,5,0.1,0.001)$, and trained for $20$ epochs. 

To test the controller, the system was initialized at equilibrium and at $t=0.2$s a three-phase ground fault occurred at B2. After $0.23$s the fault was removed and one line between B2 and B3 was tripped as a protection mechanism. As shown in Figure \ref{smib_ctrl}, the converter frequency dropped and oscillated drastically in the post-fault stage without the proposed control, i.e. with $u=0$ in the dynamics \eqref{Swing}, losing synchronization with the infinite bus, while the dissipativity-based control stabilized the voltage angle dynamics of the VSG, indicating an enlarged RoA of the post-fault equilibrium.

\subsection{Kundur Two-Area System}
\begin{figure}[!ht]
 \vspace{-2mm}
\centerline{\includegraphics[width=0.9\linewidth]{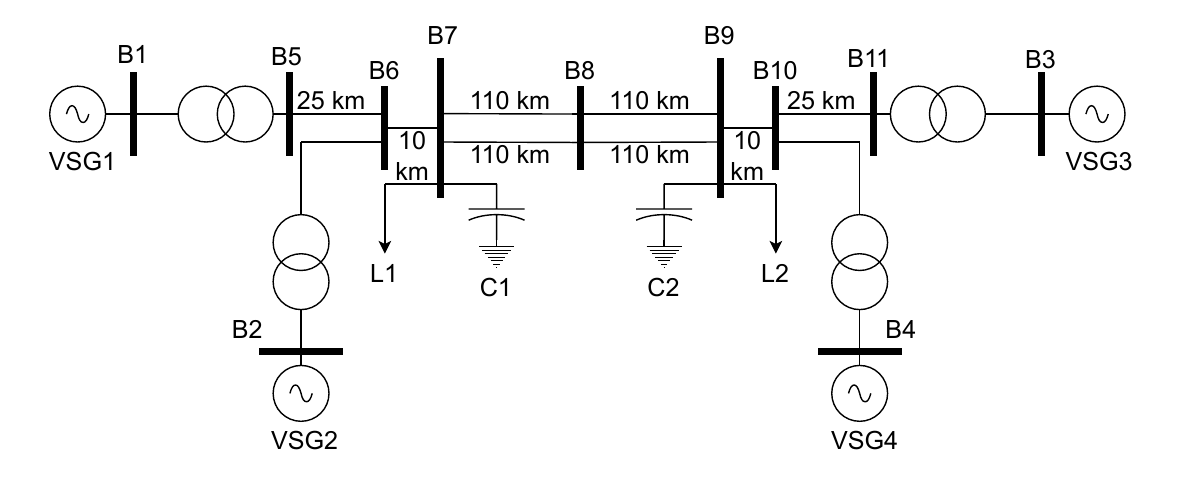}}
	\vspace{0mm}
        \caption{The modified Kundur two-area system with VSGs.}
        \label{2a}
	\vspace{-3mm}
\end{figure}
We also tested our approach on the Kundur two-area system \cite{kundur1994power} modified to include four VSGs shown in Figure \ref{2a}. The power and voltage set points of VSGs were the same as in the original Kundur system. The virtual impedance of each VSG was j$0.3$p.u., the virtual inertia was $4$s, and the damping coefficient was $5$p.u. To obtain isolated equilibrium points, we used voltage angles with respect to VSG1 and the angular frequencies as states.

To train the matrix NNs, we first used the same random sampling method as in the SCIB case to generate $2500$ trajectories of $1$s with sampling and control interval of $\Delta t =0.5$ms. The trajectories were gathered in the post fault system, where one line between B8 and B9 in Figure \ref{2a} was open. Next, we filtered the data and kept those that belong to a region of interest $\tilde{\mathcal{X}}$ satisfying $\Vert\Delta\omega\Vert_\infty\leq0.1$ and $\Vert\Delta\delta\Vert_\infty\leq2\pi$, where $\Vert\cdot\Vert_\infty$ is the infinity norm. Then we used a learning rate of $10^{-3}$ and weight decay $10^{-4}$ in Adam, $(w_1,w_2,w_3,w_4)=(1,1,0.1,10^{-3})$, and trained for $20$ epochs.

\begin{figure}[!ht]
 \vspace{-3mm}
\centerline{\includegraphics[width=0.8\linewidth]{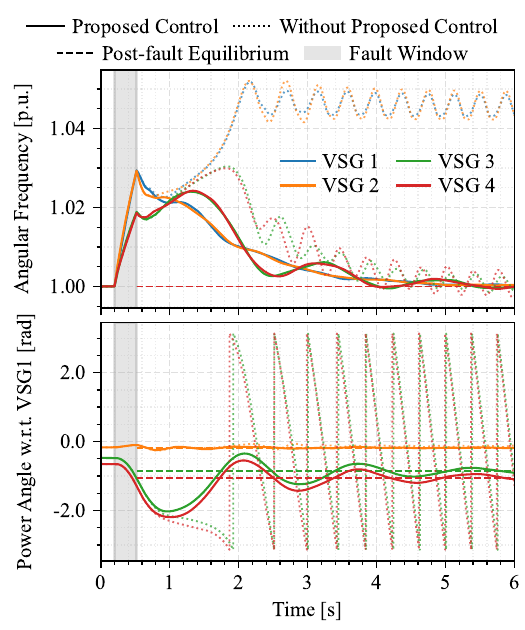}}
	\vspace{0mm}
        \caption{The effect of proposed control in the two-area system.}
        \label{2a_ctrl}
	\vspace{-3mm}
\end{figure}

To test the controller, we initiated the system at equilibrium and set a three-phase ground fault at $t=0.2$s on one circuit between B8 and B9 near B8. After $0.32$s the fault line was removed and necessary adjustments were made to the power and inner-loop set points of the VSGs to render a post-fault equilibrium with nominal frequency possible. Figure \ref{2a_ctrl} shows that without the proposed control, VSG3 and VSG4 failed to synchronize with the other area with VSG1 and VSG2, and the frequency of each VSG fluctuated greatly, while the dissipativity-based neural control stabilized the VSG-based system in the post-fault stage, suggesting an enlarged RoA of the post-fault equilibrium.

\section{Conclusions}\label{Conclusion}
We propose a novel direct data-driven method for designing stabilizing controls for VSG-based power systems with unknown dynamics. The method utilizes neural networks to learn matrices that characterize the dissipativity of the unknown system while penalizing stability condition violations. In consequence, the matrix NNs can synthesize a stabilizing feedback control with a state-dependent gain. With extra degrees of freedom, the method can also integrate cost function shaping to enhance the optimality of the controller with respect to user-defined objectives. Numerical experiments on the transient stability of VSG-based power systems demonstrate the effectiveness of our algorithm. Future work will investigate decentralized or distributed applications, conditions on data sufficiency, evaluation and improvement of robustness, and equilibrium-independent dissipativity-based neural control.

\appendices
\section{Proof of Theorem 2}\label{Apdx:T2}
\textbf{Proof}: When the control input $u$ in \eqref{QSR} is given by the control law \eqref{control}, the dissipativity condition \eqref{diss_comp} implies
\begin{equation*}
    V(x_{k+1})-V(x_k)\leq-x_k^T\Phi(x_k)x_k<0,\forall x_k\in\mathcal{X}/\{0\},
\end{equation*}
where we have used condition \eqref{delta} to rewrite the supply rate under the feedback \eqref{control}.

Since $V(x)$ is positive definite, it becomes a Lyapunov function as \eqref{Lyapunov} requires. Thus, $\pi(x)$ is a stabilizing control.\qed

\section{Proof of Theorem 3}\label{Apdx:T3}
\textbf{Proof}: According to Theorem 2, the asymptotical stability condition is satisfied. Condition \eqref{diss_comp} implies that $\tilde l(x_k,u_k)$ is nonnegative, while condition \eqref{delta} indicates the positive definiteness of $x_k^T\Phi(x_k)x_k$. Consider the optimal cost:
\begin{equation*}
\begin{aligned}
J =& \min_{\{u_k\}_{k=0}^\infty}\sum_{k=0}^{\infty}\left[\tilde l(x_k,u_k)+x_k^T\Phi(x_k)x_k\right]\\
=&-\sum_{k=0}^{\infty}\left[V(x_{k+1})-V(x_{k})\right]\\
&+\min_{\{u_k\}_{k=0}^\infty}\sum_{k=0}^{\infty}\left[s (x_k,u_k)+x_k^T\Phi(x_k)x_k\right].\\
\end{aligned}
\end{equation*}

As \eqref{control} is a stabilizing control, the first summation reduces to $-\sum_{k=0}^{\infty}\left[V(x_{k+1})-V(x_{k})\right]=V(x_0)-\lim_{k\to\infty}V(x_k)=V(x_0)$. In the second summation, for each fixed $x_k$, $s(x_k,u_k)$ is quadratic in $u_k$. Since the quadratic coefficient $R(x_k)$ is positive definite, $s(x_k,u_k)$ attains its minimum value $-x_k^T\Phi(x_k)x_k$ when $u_k$ follows the feedback law \eqref{control}. Therefore, the second summation $\sum_{k=0}^{\infty}\left[s (x_k,u_k)+x_k^T\Phi(x_k)x_k\right]$ attains its minimum $0$ under the same feedback law. In conclusion, the control law \eqref{control} is optimal with the minimal cost $J=V(x_0).$ \qed


\end{document}